%% file: FEpaper.tex
\documentclass[prb,twocolumn,showpacs,showkeys]{revtex4}

\usepackage{graphicx}
\usepackage{color}
\usepackage{amssymb}
\usepackage{amsmath}

\newcommand{\ket}[1]{|#1\rangle}
\newcommand{\bra}[1]{\langle #1|}
\newcommand{\expct}[2]{\left\langle #1 \right\rangle_{#2}}
\newcommand{\expcts}[2]{\langle #1 \rangle_{#2}}
\newcommand{\sgn}{\text{sgn}}

\begin{document}

\title{On the visibility of electron-electron\\ interaction effects in field emission spectra}

\author{T. L. Schmidt and A. Komnik}

\affiliation{Physikalisches Institut,
 Albert--Ludwigs--Universit\"at, D--79104 Freiburg, Germany}

\begin{abstract}
One of the most convenient methods to obtain information about the
energy distribution function of electrons in conducting materials
is the measurement of the energy resolved current $j(\omega)$ in
field emission (FE) experiments. Its high energy tail
$j_>(\omega)$ (above the Fermi edge) contains invaluable
information about the nature of the electron--electron
interactions inside the emitter. Thus far, $j_>(\omega)$ has been
calculated to second order in the tunnelling probability, and it
turns out to be divergent toward the Fermi edge for a wide variety
of emitters. The extraction of the correlation properties from
real experiments can potentially be obscured by the eventually
more divergent contributions of higher orders as well as by
thermal smearing around $E_F$. We present an analysis of both
factors and make predictions for the energy window where only the
second order tunnelling events dominate the behaviour of
$j_>(\omega)$. We apply our results to the FE from Luttinger
liquids and single-wall carbon nanotubes.
\end{abstract}



\maketitle

\section{Introduction}
According to the original idea of Fowler and Nordheim \cite{fn},
the cold emission of electrons from metallic electrodes can be
considered as tunnelling of particles from the conduction band
into the vacuum through a triangular barrier. The precise form of
the latter is determined by the work function $W_A$, by the
applied electric field $F$ as well as by the geometry of the
emitter tip, and is characterised by the effective transmission
coefficient $D(\omega)$ for the electrons with energy $\omega$
\cite{gadzuk_plummer}. The most important measurable quantities
are the energy resolved current $j(\omega)$ and the total current
$J=\int d\omega \, j(\omega)$. The currents drawn from the
emitters are usually very small so that $j(\omega)$ is
proportional to the equilibrium energy distribution function of
electrons in the vicinity of the tip, $n_F(\omega)$, and to
$D(\omega)$ \cite{fn},
\begin{equation}
  j(\omega) \sim D(\omega) n_F(\omega).
\end{equation}
Therefore, at $T=0$ no electrons are allowed with energies above
the Fermi edge $E_F$. However, in 1970, Lea and Gomer succeeded in
measuring $j(\omega)$ to a very high degree of accuracy. They
found significant contributions above $E_F$ \cite{lea-gomer}. This
effect can be attributed to electron-electron interactions inside
the emitter. The relevant process is usually referred to as
``secondary tunnelling'' \cite{gadzuk_plummer}, during which an
electron with energy below the Fermi edge tunnels out of the
emitter leaving behind a hot hole. In the second stage, this hole
scatters inelastically, producing a secondary electron which can
obtain an energy above $E_F$. Due to its higher energy, this
electron has an increased tunnelling probability and will
contribute to a ``secondary current'' $j_>(\omega)$ above $E_F$.
Since $j_>(\omega)$ comes about as a result of interactions, its
measurement reveals valuable information about the nature of
correlations inside the emitter.

Two different types of rigorous calculations of $j_>(\omega)$ have
been attempted so far. The approach of \cite{gadzuk_plummer1} used
kinetic equations. In \cite{komnik1} the high-energy tails in
$j(\omega)$ were calculated using non-equilibrium perturbation
theory in the tunnelling probability. One of the principal
predictions is the asymptotic behaviour toward the Fermi edge,
which turns out to be divergent. Both approaches take into account
only tunnelling processes of lowest order. To the best of our
knowledge thus far no investigations of processes of higher orders
have been undertaken despite the fact that they can potentially be
even more divergent around $E_F$ and therefore obscure the
predicted effects. With this work we would like to address this
problem and analyse the influence of higher order tunnelling
processes on $j_>(\omega)$.

 Thermal broadening of the Fermi edge can also prevail
over the contributions of lowest order, making their measurement
difficult if not impossible. While in conventional Fermi liquid
systems this broadening is rather simple to estimate, in many
interacting systems it has not yet been analysed. One of the most
important of these is the universality class of 1D correlated
electrons or the so-called Luttinger liquids (LLs), which are e.g.
realised in single-wall carbon nanotubes (SWNTs), as has been
shown both theoretically and experimentally
\cite{egger_gogolin_prl,kane_fisher,bockrath,yao}.

\begin{figure}[ht]
\centering
\input{aufbau.pstex_t}
\caption{Schematic view of the system under consideration}
\label{fig:fielde}
\end{figure}
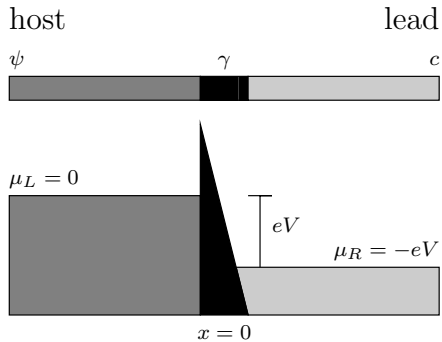
Recently, the advancement in manufacturing enabled the
construction of the first flat panel display which employs carbon
nanotubes as emitters \cite{fan_fed,choi_fed}. Refinement and
further development of this technology requires detailed knowledge
of their emission spectra. This is the reason why it is important
to fully understand the field emission properties of SWNTs. Thus
far, theoretical activities in this field have been rather
moderate.
In this work, we would like to analyse the effects of tunnelling
processes of higher orders from LLs along with the effects of
thermal broadening of the Fermi edge on the high-energy tail of
$j_>(\omega)$ and apply the results to FE from SWNTs.




\section{Next-to-leading order in tunnelling: diagrammatic approach}
\label{principalpart}
We model the field emission set-up by a heterogeneous tunnelling
junction (Fig.~\ref{fig:fielde}). The left contact, which will be
referred to as ``host'', represents the emitter. The chemical
potential of this system is set to $\mu_L = 0$. We shall assume
that tunnelling only takes place precisely at the interface
between the conductors. The right contact will be referred to as
``lead'', and is represented by a non-interacting Fermi gas with
chemical potential $\mu_R = -eV$. As during field emission
electrons tunnel into vacuum, this potential plays the role of a
cut-off and has to be set to the conduction band width or the
characteristic energy scale entering $D(\omega)$, whichever is
larger. The transmission coefficient $D(\omega)$ of the triangular
barrier encountered in field emission depends exponentially on the
energy and the work function $W_A$. We restrict ourselves to the
low temperature limit, where interesting processes will only
involve electrons quite close to the Fermi edge. That is why we
replace this tunnelling probability by a constant, $\gamma^2 \sim
D(0)$, where $\gamma$ is the energy independent tunnelling
amplitude (we set $E_F=0$ throughout). The model Hamiltonian is
then given by
\begin{equation}\label{eq:Hamiltonian}
H = H[\psi] + H[c] + \gamma[\psi^\dag(0)c(0) + \psi(0)c^\dag(0)],
\end{equation}
consisting of three parts. $H[\psi]$ and $H[c]$ are the
Hamiltonians of the host and the lead, respectively, and the third
term describes the tunnelling events.

Due to the finite applied voltage, the system under consideration
is in a non-equilibrium state. The most convenient method of
describing such systems is the Keldysh formalism \cite{keldysh},
where we shall be taking $H_{int} = \gamma[\psi^\dag(0)c(0) +
\psi(0)c^\dag(0)]$ as the perturbation of the Hamiltonian $H_0 =
H[\psi] + H[c]$. We shall use the notation employed in \cite{ll},
where the Green's functions are defined by
\begin{eqnarray}
  g^{--}(t, t') & = & -i\expcts{T[\psi(t) \psi^\dag(t') S_C]}{0}, \nonumber \\
  g^{-+}(t, t') & = &  i\expcts{\psi^\dag(t')\psi(t) S_C}{0}, \nonumber \\
  g^{+-}(t, t') & = & -i\expcts{\psi(t) \psi^\dag(t') S_C}{0}, \nonumber \\
  g^{++}(t, t') & = & -i\expcts{\tilde{T}[\psi(t)
  \psi^\dag(t') S_C]}{0} \, .
\end{eqnarray}
$T$ and $\tilde{T}$ are time and anti-time ordering operators,
respectively, and $\expcts{\cdots}{0}$ stands for the averaging
with respect to the ground state of $H_0$. We use $g(t, t')$ to
denote Green's functions of particles in the host, whereas $G(t,
t')$ refers to Green's functions in the lead. The operator $S_C$
is defined by the perturbation,
\begin{equation}\label{eq:s_matrix_fe}
  S_C = T_C \exp \left\{ - \frac{i\gamma}{\hbar} \int_C dt
  \left[\psi^\dag(t)c(t) + \psi(t)c^\dag(t) \right] \right\} \, .
\end{equation}
As shown in \cite{komnik}, the energy resolved current for a
certain energy $\omega > 0$ is
proportional to the local Green's function $g^{-+}(\omega)$ in the
host. Our primary task is to calculate the function
$g^{-+}(\omega)$. For this purpose we shall perform a perturbative
expansion of $S_C$ in the tunnelling amplitude $\gamma$, so that
$g^{-+}(\omega) = g_0^{-+}(\omega) + g_2^{-+}(\omega) +
g_4^{-+}(\omega) + \ldots$ where $g_2 \sim \gamma^2$ and $g_4 \sim
\gamma^4$. As the Fermi edge is still sharp even in interacting
systems, the zeroth order does not contribute above $E_F$.
According to \cite{komnik1}, the second order correction to the
Green's function can be brought to the form,
\begin{eqnarray}\label{eq:g2}
  g^{-+}_2(\omega) & = & \frac{\gamma^2}{\hbar^2} \int dt e^{i\omega t/\hbar}
  \iint_{-\infty}^\infty dt_1 dt_2\\
  & \times &  \sum_{ij = \pm}
  (ij) K_{ij}(t, t_1, t_2) G^{ij}(t_1, t_2) \nonumber,
\end{eqnarray}
where $K_{ij}$ are four-point correlators of $\psi$-operators
which correspond to rectangular diagrams with
insertions\cite{komnik}. It has been shown in \cite{komnik} that
only
\begin{eqnarray}\label{eq:4corr}
    K_{+-}(t, t_1, t_2) & = &
    - \langle \tilde{T} \{ \psi^\dag(0) \psi^\dag(t_1) \} T\{ \psi(t) \psi(t_2) \} \rangle \label{eq:4corr_last}
\end{eqnarray}
yields a non-zero contribution above $\omega = 0$, which diverges
toward the Fermi edge.

\begin{figure}
\centering
\includegraphics[width=4cm]{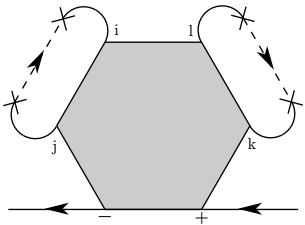}
\caption{{\bf Fourth order diagram:} The hexagon corresponds to a
six-point correlation function of $\psi$ operators and the indices
$i,j,k,l = \pm$ denote the respective part of the Keldysh contour.
Solid lines depict Green's functions in the host, dashed lines
stand for Green's function is the lead wire. Crosses stand for the
tunnelling vertices.}\label{fig:fourth_order_corr}
\end{figure}

We can use the same approach that led to (\ref{eq:g2}) and expand
to the next-to-leading order in $\gamma$. It turns out to be given
by
\begin{eqnarray}\label{eq:4th_order_final}
  g^{-+}_4(\omega) & = & \frac{i\gamma^4}{2\hbar^4} \int dt e^{i\omega t/\hbar}
  \iiiint_{-\infty}^\infty dt_1 \cdots dt_4  \\
  & \times & \sum_{ijkl} K_{ijkl}(t, t_1, t_2, t_3, t_4)
  G^{ij}(t_1, t_2) G^{kl}(t_3, t_4) \, , \nonumber
\end{eqnarray}
which schematically corresponds to a sum over 16 different
hexagon-shaped diagrams, s. Fig.~\ref{fig:fourth_order_corr}, each
containing \emph{two} insertions of the lead Green's functions and
a six-point correlator $K_{ijkl}$ which is similar to the
four-point correlators in the second-order case. The structure of
the diagrams is not surprising as in this order two successive
tunnelling processes are allowed. Since we are only interested in
the high-energy contributions above $E_F$, we first have to
identify the diagrams which may contribute in this situation. As
in the case of lowest order contributions, the best
way to identify the relevant constellations without explicit
calculation is to take advantage of the spectral representation.
To that end we have to be able to perform the $t$-integration in
(\ref{eq:4th_order_final}). This can be accomplished by insertion
of an appropriate number of complete sets of eigenstates, thereby
extracting the $t$-dependence.

In order to illustrate the procedure, we shall determine the
contribution of the $K_{----}$ correlation function where the time
indices $t_1$ to $t_4$ are taken on the time-ordered ``$-$''
shoulder of the Keldysh contour. Hence, we wish to calculate
\begin{eqnarray}\label{eq:g---}
  g^{-+}_4(\omega) & = &\frac{i\gamma^4}{4\hbar^4} \int dt e^{i\omega t/\hbar}
  \iiiint dt_1 \cdots dt_4 \\
& \times &
  \expct{\psi^\dag(-t) T[\psi(0) \psi^\dag_1 \psi_2 \psi^\dag_3 \psi_4]}{0}
  \expct{T[c_1 c^\dag_2 c_3 c^\dag_4]}{0}. \nonumber
\end{eqnarray}
The time development of the $\psi(t)$ and $c(t)$ operators in the
interaction picture is governed by the unperturbed Hamiltonians
$H[\psi]$ and $H[c]$, respectively. Although due to the
interactions in the host, we cannot diagonalise $H[\psi]$, we
still know that there is a complete set of eigenstates satisfying
$H[\psi]\ket{\nu} = E_\nu \ket{\nu}$ and $\sum_{\nu}
\ket{\nu}\bra{\nu} = 1$ with positive eigenenergies $E_\nu$.
Inserting a complete set into (\ref{eq:g---}) yields
\begin{eqnarray}
  g^{-+}_4(\omega) & = &\frac{i\gamma^4}{4\hbar^4} \sum_\nu \int dt e^{i\omega
  t/\hbar}\\
  & \times & \bra{0} e^{-i H[\psi] t/\hbar} \psi^\dag e^{i H[\psi] t/\hbar} \ket{\nu}
  A(\nu),\nonumber
\end{eqnarray}
where $A(\nu)$ is independent of $t$,
\begin{eqnarray}
  A(\nu) & = & \iiiint dt_1 \cdots dt_4 \\
  & \times & \bra{\nu} T[\psi(0) \psi^\dag_1 \psi_2 \psi^\dag_3
  \psi_4]\ket{0}\expcts{T[c_1 c^\dag_2 c_3 c^\dag_4]}{0}.\nonumber
\end{eqnarray}
Hence, the $t$ integration can be performed and leads to
\begin{eqnarray*}
  g^{-+}_4(\omega) & = &\frac{i\gamma^4}{4\hbar^4} \sum_\nu \int dt e^{i\omega t/\hbar}
  e^{i E_\nu t/\hbar} \bra{0} \psi^\dag \ket{\nu} A(\nu) \\
  & = & \frac{i\gamma^4}{4\hbar^3} \sum_\nu \delta(\omega + E_\nu)
  \bra{0} \psi^\dag \ket{\nu} A(\nu).
\end{eqnarray*}

As all energies $E_\nu$ are positive, we conclude that this
function does not yield a contribution above $\omega = 0$. Similar
calculations can be performed for all other correlation functions.
We find out, that e.g. in second order, the only contributing
diagram is described by the correlation function,
\begin{eqnarray}
    K_{+-} & = &
    - \langle \tilde{T} [ \psi^\dag(0) \psi^\dag_1 ] T[ \psi(t) \psi_2]
    \rangle,
\end{eqnarray}
which contributes for $\omega < eV$, the upper threshold for
second order processes. On the other hand, in fourth order there
are only three relevant contributions which are given by
\begin{eqnarray}
  K_{+---} & = & \expct{\tilde{T}[\psi^\dag(0) \psi^\dag_1]
  T[\psi(t) \psi_2 \psi^\dag_3 \psi_4]}{0}, \\
  K_{+-+-} & = & \expct{\tilde{T}[\psi^\dag(0) \psi^\dag_1 \psi^\dag_3]
  T[\psi(t) \psi_2 \psi_4]}{0}, \\
  K_{+++-} & = & \expct{\tilde{T}[\psi^\dag(0) \psi^\dag_1 \psi_2 \psi^\dag_3]
  T[\psi(t) \psi_4]}{0}.
\end{eqnarray}
Two of these correlation functions ($K_{+---}$ and $K_{+++-}$)
contribute up to $\omega = eV$ and have to be counted twice due to
reasons of symmetry. The correlation function $K_{+-+-}$
contributes for $\omega < 2eV$ which is the upper threshold for
fourth order processes.
In all of these contributions at least one of the insertions is of
the type encountered in the lowest order. We know that in the
lowest order (the rectangular diagram of \cite{komnik}) the purely
time or anti-time ordered insertions do not lead to contributions
above $\omega=0$. Therefore, in the terms where only one of the
insertions is of the type $G^{+-}(\omega)$, one can regard the
corresponding diagram as an effectively rectangular one with
respect to high-energy tails.

\section{Field emission from Luttinger liquids and SWNTs.}
\label{applications}
The usual way to proceed after the relevant diagrams have been
identified is to perform a perturbative expansion of all
contributions in the interaction constant. However, a perturbative
breakdown of the six-point correlation functions (hexagons of
Fig.~\ref{fig:fourth_order_corr}) in non-equilibrium involves a
very large set of diagrams. Performing the same calculation for a
LL system does not have that disadvantage. The reason for that is
the fact that in the bosonization representation of an LL any
possible fermionic correlation function can be computed for any
interaction strength  without resorting to approximation schemes
\cite{luttinger,tomonaga,haldane}. The perturbative results are
then expected to be recovered after taking the limit of vanishing
interactions. On the other hand, as we are showing later, the
results for an LL only have to be slightly modified in order to be
applied to the experimentally relevant FE from SWNTs.

Our main goal is the calculation of the kernels $K_{ijkl}$. We
assume the emitter to be infinite extending from $x = -\infty$ to
$x=0$. Hence, we are dealing with a semi-infinite wire with an
open boundary condition (OBC) at its end. Using OBC bosonization
\cite{fg}, exactly at the emitter tip ($x=0$), where all
tunnelling events take place, we have for the electron field
operator
\begin{equation}\label{eq:luttinger_final_psi}
  \psi(x=0,t) = \frac{1}{\sqrt{2\pi a}} e^{2\pi
  i\phi(x=0,t)/\sqrt{g}},
\end{equation}
where $a$ is an infinitesimal cut-off length which can be
interpreted as the lattice constant of the underlying lattice
model. We omit the ``ladder operator'' (particle number changing
operator) as we are later going to the limit of infinitely long
wire $L \rightarrow \infty$ \cite{haldane,grabert}. The bosonic
phase field $\phi$ entering (\ref{eq:luttinger_final_psi}) is
given by
\begin{equation}\label{eq:luttinger_final_phi}
  \phi(x=0,t) = \frac{1}{2\pi}\sum_{q > 0} \sqrt{\frac{\pi}{Lq}} \left( e^{-iqvt}a_q +
  e^{iqvt}a^\dag_q \right) e^{-aq/2}.
\end{equation}
In the above equations, $g$ is a dimensionless quantity measuring
the interaction strength \cite{haldane},
\begin{equation}
  g =  \left(1+\frac{U_0}{\pi \hbar v_F}
  \right)^{-1/2},
\end{equation}
where $U_0$ is the Fourier transform of the screened
electron-electron interaction potential $U(x-y)$ at $q = 0$ and
$v_F$ is the Fermi velocity, $v = v_F / g$.

By using the Baker-Hausdorff formula, we can calculate arbitrary
correlation functions of the type
\begin{equation}\label{eq:korr_def}
  \mathbb{K} = \langle \alpha_1 \cdots \alpha_N \rangle,
\end{equation}
where $N$ is an even number and $\alpha_i$ are fermion creation or
annihilation operators,
$\psi^\dag(t_i)$ or $\psi(t_i)$. We arrive at the result,
\begin{eqnarray}
    \mathbb{K}
    & = & \frac{F(0)^{N/2g}}{\sqrt{(2\pi a)^N}} \left[ \prod_{m=1}^{N-1} \prod_{n=m+1}^{N}
    F(t_m - t_n)^{\beta_m \beta_n} \right]^{1/g},
    \label{eq:korr_allg}
\end{eqnarray}
where
\begin{equation}
  F(t) = 1-e^{-i \epsilon_0 \left( t - i \delta \right)}.
\end{equation}
The parameters $\epsilon_0$ and $\delta$ are defined by
$\epsilon_0 = \pi/L v$ and $\delta = a/v$. Furthermore $\beta_i =
+1$ [$\beta_i = -1$] for $\alpha_i = \psi(t_i)$ [$\alpha_i =
\psi^\dag(t_i)$].

Using (\ref{eq:korr_allg}) and taking into account the time
orderings, we obtain for the rectangular diagram \cite{komnik}
\begin{eqnarray}\label{eq:g2_F}
  g^{-+}_2(\omega) & = & -\frac{\gamma^2}{(2\pi a \hbar)^2} \int dt e^{i\omega
  t/\hbar}  \iint_{-\infty}^\infty dt_1 dt_2 \nonumber \\
  & \times & \sgn(-t_1) \sgn(t - t_2) G^{+-}(t_1, t_2) \nonumber \\
  & \times &
    \left[ \frac{F(0)^2 F(-|t_1|) F(|t - t_2|)}{F(-t) F(-t_2) F(t_1 - t) F(t_1
  - t_2)}  \right]^{1/g}.
\end{eqnarray}
Assuming the lead to be a Fermi gas with chemical potential $\mu_R
= -eV$, the corresponding Green's function becomes
\begin{eqnarray}
  G^{+-}(t_1, t_2)
  & = & -\frac{\hbar\nu}{2\pi} \frac{e^{ieV(t_1-t_2)/\hbar}}{t_1-t_2-i \delta} \label{eq:G_0},
\end{eqnarray}
with an infinitesimal parameter $\delta$. For reasons of
simplicity, we have assumed a constant density of states $\nu$ in
the lead. Furthermore, we approximate $F(t)$ by $F(t) \approx
i\epsilon_0(t-i\delta)$, which is justified for $L \rightarrow
\infty$. These approximations allow the calculation of
$g^{-+}_2(\omega)$, which has been performed in \cite{komnik}.
However, the approach of \cite{komnik} cannot be adapted to the
fourth order. Nevertheless, as $\gamma$ is very small,
$g^{-+}_4(\omega)$ can become important only in the immediate
vicinity of the Fermi energy and even that only when it turns out
to be more divergent than the lowest order term. Therefore, it is
sufficient to find the $\omega \rightarrow 0$ asymptotics and we
require $eV/\omega \rightarrow \infty$. In that case, the Green's
function $G^{+-}$ can be shown to become a $\delta$-function,
\begin{equation}\label{eq:g_delta}
  G^{+-}(t_1, t_2) \rightarrow  -i\hbar \nu \delta(t_1 - t_2).
\end{equation}
Using this representation, the high energy particle number
$n_2(\omega) = -ig^{-+}_2(\omega)/\hbar$ can readily be calculated
and one immediately obtains
\begin{equation}\label{eq:final_2nd_order}
  n_2(\omega) = n^c_2 \left(\frac{\omega}{\omega_c}\right)^{1/g-2},
\end{equation}
where $\omega_c = \gamma/a$ and
\begin{equation}\label{eq:n2c}
  n^c_2 = \frac{\nu}{(2\pi)^2} \left(\frac{\gamma}{v\hbar}\right)^{1/g}
  k_{2}.
\end{equation}
$k_2$ contains a numerical prefactor which is given by the
dimensionless convergent integral
\begin{eqnarray}\label{eq:k2}
  k_2 & = &
  (-i)^{1/g}\lim_{\delta\rightarrow 0}\int dy e^{i y} \int_{-\infty}^\infty dy_1
  \sgn(-y_1)\sgn(y-y_1) \nonumber \\
  & \times & \left[
    \frac{(-|y_1|)(|y-y_1|)}{(-y-i\delta)(-y_1-i\delta)(y_1-y-i\delta)}
    \right]^{1/g}.
\end{eqnarray}
Within the approximation $eV/\omega \rightarrow \infty$, the
calculation of the fourth order contributions can be performed
likewise. We start with the determination of the contribution of
the $K_{+-+-}$ correlator. In analogy with (\ref{eq:g2_F}), we
obtain a similar expression but with two insertions. According to
(\ref{eq:g_delta}) we obtain two $\delta$ functions which allow to
perform two of the four integrations. Switching to dimensionless
quantities, we obtain
\begin{equation}
  g^{-+}_4(\omega) =
  \frac{i\hbar \nu}{(2\pi)^3}
  \frac{\nu\gamma}{2} k'_4
  \left(\frac{\gamma}{v\hbar}\right)^{1/g}
  \left(\frac{\gamma}{a\omega}\right)^{3-1/g} \, ,
\end{equation}
where the numerical factor $k'_4$ is given by
\begin{eqnarray}\label{eq:k4prime}
  k'_4 & = & (-i)^{1/g}\int dy e^{iy} \iint dy_1 dy_3 \\
  & \times & \frac{\sgn(-y_1) \sgn(-y_3) \sgn(y-y_1)\sgn(y-y_3) }{(-y-i\tilde{\delta})^{1/g}}
  \nonumber \\
  & \times &
  \left[ \frac{-|y_1| |y_3| |y_1 - y_3|}
  {(-y_1-i\tilde{\delta})(-y_3-i\tilde{\delta})(y_1-y-i\tilde{\delta})}
   \right]^{1/g} \nonumber \\
  & \times &
  \left[ \frac{ |y-y_1| |y-y_3| |y_1-y_3| }
  {(y_1-y_3-i\tilde{\delta})(y_3-y-i\tilde{\delta})(y_3-y_1-i\tilde{\delta})}
  \right]^{1/g}.\nonumber
\end{eqnarray}
The other diagrams, $(ijkl) = (+---)$ and $(ijkl) = (+++-)$, can
be treated in the same way with the exception that the time
ordering combinatorics is significantly more complicated.
However, we found that all three diagrams yield the same leading
power-law energy dependence,
\begin{equation}\label{eq:final_4th_order}
  n_4(\omega) = n^c_4
  \left(\frac{\omega}{\omega_c}\right)^{1/g-3},
\end{equation}
with
\begin{equation}
  n^c_4 = \frac{\nu}{(2\pi)^3} \frac{\nu \gamma}{2} \left(
  \frac{\gamma}{v\hbar}\right)^{1/g} k_4.
\end{equation}
For the $(ijkl) = (+-+-)$ case the corresponding $k_4'$ is given
by (\ref{eq:k4prime}). Numerical evaluation of $k_4'$ yields a
value around unity. Due to the different time ordering
combinatorics of the six-point correlation functions, the
numerical prefactors $k_4''$ and $k_4'''$ for the other two
diagrams cannot be written in a compact form similar to
(\ref{eq:k4prime}).  As the other two diagrams are related to
those of the second order in tunnelling (see Section
\ref{principalpart}), $k_4''$ and $k_4'''$ have to be of the same
order of magnitude as $k_4'$.

Equation (\ref{eq:final_4th_order}) is the central result of this
paper. The contribution of next-to-leading order processes turns
out to be even more divergent than the leading order in case of
weak interactions, $g>1/2$. However, we expect the overall
prefactor to be negative as in the fourth order in tunnelling the
recombination of the secondary holes and secondary electrons,
stemming from different primary processes, is allowed, effectively
suppressing the high-energy tails in $j_>(\omega)$. Taking the
limit $g \rightarrow 1$ we expect to obtain the asymptotics of the
next-to-leading term for a Fermi liquid, $n_4(\omega) \sim
1/\omega^2$. Similarly to the leading order tunnelling results, at
$g=1$ $j_>(\omega)$ is identically zero due to a vanishing
prefactor $k_4$. That can be shown analytically.

So far, all calculations have been made for the case of a spinless
Luttinger liquid. However, carbon nanotubes are known to be
described in terms of a four-channel LL
\cite{egger_gogolin,egger_gogolin_prl,kane_fisher}. The four
channels are responsible for the charge-flavour ($\phi_{c-}$),
total spin ($\phi_{s-}$), spin-flavour ($\phi_{s+}$) and total
charge ($\phi_{c+}$) excitations. The three former are
non-interacting ($g=1$) whereas the latter possesses the Luttinger
parameter $g_{c+}=(1+4\tilde{V}_0/\pi \hbar v_F)^{-1/2}$, where
$\tilde{V}_0$ is the $k=0$ component of the Fourier transform of
the screened interaction potential. It can further be shown that
the field operator $\psi$ can be written in terms of Bose fields
as
\begin{equation}
  \psi = \frac{1}{2\pi a}\exp\left[\pi i (\phi_{c+}/\sqrt{g_{c+}}
  + \phi_{c-} + \phi_{s+} + \phi_{s-}) \right].
\end{equation}
As different species of $\phi$ operators commute, all correlation
functions factorise. Therefore, the results for the one-channel LL
can be applied to SWNTs by the simple replacement
\begin{equation}
  g^{-1} \rightarrow \frac{g_{c+}^{-1} + 3}{4}.
\end{equation}
An estimation of the typical interaction parameter $g$ can be done
using the following formula, which relates the geometry of the
nanotube to the interaction constant $g_{c+}$
\cite{egger_gogolin},
\begin{equation}
  g_{c+} = \left(1 + \frac{8e^2}{\pi \kappa \hbar
  v_F}\left[\ln(L/2\pi R) + 0.51\right] \right)^{-1/2},
\end{equation}
where $\kappa$ is the dielectric constant, $L$ is the length of
the nanotube and $R$ its radius. Assuming typical values we arrive
at an interaction constant $g_{c+} \approx 0.2$ which leads to an
effective $g \approx 0.5$.

\section{Visibility of high energy tails}
\label{visibility}
Thus far, we have discussed the zero temperature case, where a
non-trivial high-energy tail in the energy resolved current is
produced by the interaction effects. However, such a tail can
trivially emerge as a result of the thermal broadening of the
Fermi edge. In this Section, we shall address the problem of
distinguishing the correlation effects from those of finite
temperature restricting ourselves to the model of spinless LL.

We first calculate the width of the broadening as a function of
interaction constant and temperature. Obviously, the main
contribution to $j_>(\omega)$ simply stems from the zeroth order
particle density for a Luttinger liquid, which can be brought to
the form \cite{delft}
\begin{equation}
  n_T(\omega) = \frac{1}{h a} \int_{-\infty}^{\infty} dt e^{i \omega t/\hbar}
  \frac{\sin^\lambda\left(\frac{\pi
  a}{\hbar\beta v}\right)}{\sin^\lambda\left[\frac{\pi}{\hbar\beta v}(-i v t +
  a)\right]},
\end{equation}
where $\lambda = 1/g$. In order to perform this integral, is is
convenient to express the sine function in the denominator as an
exponential and to rewrite the exponent $\lambda$ in the
denominator with the help of the binomial formula. Then, the
integral can be performed and we obtain a representation of
$n(\omega)$ in terms of the hypergeometric function $F$
\cite{abramowitz},
\begin{equation}
  n_T(\omega) = \frac{\sin^\lambda(\frac{\pi a}{\hbar\beta v}) (2
  i)^\lambda}{h a} \left[ I_\lambda(\omega) + (-1)^\lambda
  I^*_\lambda(\omega)
  \right]
\end{equation}
with
\begin{eqnarray}\label{eq:I_hypergeom}
  I_\lambda(\omega) & = & \frac{\hbar \beta}{\pi \lambda -
  i\beta\omega}\\
  & \times & F\left(\lambda, \frac{\lambda}{2} -
  \frac{i\beta\omega}{2\pi},1 + \frac{\lambda}{2} -
  \frac{i\beta\omega}{2\pi},
  e^{-2\pi i a/(\hbar \beta v)}\right) \nonumber .
\end{eqnarray}
This function only depends on the combination $\beta\omega$ rather
than on $\omega$ alone. Consequently, we can deduce that the width
of the thermal broadening will be of the form $\omega_C = k_B T
f(\lambda)$ with a function $f(\lambda)$ independent of the
temperature.

Using an appropriate expression for the hypergeometric function in
the case $\lambda = 1$ and using the expansion for $a \rightarrow
0$, the Fermi distribution function of the non-interacting case
can easily be recovered. For the interacting case, i.e. for
$\lambda
> 1$, we need the following expansion \cite{abramowitz},
\begin{eqnarray}
  F(x, y, z, e^{i\varphi}) & \approx &
  \frac{\Gamma(z)\Gamma(x+y-z)}{\Gamma(x)\Gamma(y)} \\
  & \times & (1-e^{i\varphi})^{z-x-y}\left[ 1 + O(e^{i\varphi} - 1)\right] \nonumber \\
  & + & \frac{\Gamma(z)\Gamma(z-x-y)}{\Gamma(z-x)\Gamma(z-y)}
  \left[ 1 + O(e^{i\varphi} - 1)\right]\nonumber .
\end{eqnarray}
Neglecting terms of order $O(e^{i\varphi} -1)$ which is
appropriate in the limit of small $a$, after some algebraic
transformations we arrive at
\begin{eqnarray}\label{eq:nT}
  n_T(\omega) & = & \frac{1}{\hbar v} \left(\frac{2\pi a}{\hbar
  \beta v}\right)^{\lambda-1} \Gamma(1-\lambda) \sin(\pi\lambda)
  \\
  & \times &
  \frac{e^{-\beta\omega/2}  }{[\cosh(\beta\omega) -
  \cos(\pi\lambda)] \left|
  \Gamma(1-\frac{\lambda}{2}-\frac{i\beta\omega}{2\pi})
  \right|^2}\nonumber
\nonumber.
\end{eqnarray}

Having established this relation for finite temperatures, we can
estimate the visibility of the second order processes by comparing
its contribution $n_2(\omega)$ to the contribution $n_T(\omega)$.
The result is
\begin{eqnarray}              \label{n2nT}
  \frac{n_2(\omega)}{n_T(\omega)} & = &\frac{1}{(2\pi)^{1/g+1}}
  \left(\frac{\nu \gamma^2 \beta}{a}\right) \frac{k_2[\cosh(\beta\omega) - \cos(\pi/g)]
  }{\Gamma(1-1/g)
  \sin(\pi/g)} \nonumber \\
  & \times &
  (\beta\omega)^{1/g-2} \left|
  \Gamma(1-\frac{\lambda}{2}-\frac{i\beta\omega}{2\pi}) \right|^2
  e^{\beta\omega/2}.
\end{eqnarray}
In order to eliminate the tunnelling amplitude $\gamma$ we can use
the following relation with the total current $I$ and the voltage
$V$ \cite{gadzuk_plummer,diplomarbeit},
\begin{equation}
  I =   g \Gamma(1-1/g)\sin(\pi/g) \frac{e \gamma^2 \nu}{2 \pi^2 \hbar a}
  \left(\frac{a e V}{\hbar v}\right)^{1/g}.
\end{equation}
Using this equation, we arrive at the following result
\begin{equation}
  \frac{n_2(\omega)}{n_T(\omega)} = \kappa f_0(\beta, I, V) f_1(\beta\omega),
\end{equation}
where
\begin{eqnarray}
\kappa & = & \frac{(2\pi)^{1-1/g} k_2}{2g\Gamma^2(1-1/g)
    \sin^2(\pi/g)},\\
f_0(\beta, I, V) & = & \frac{\hbar \beta I}{e} \left(\frac{\hbar
    v}{a e V}\right)^{1/g}, \\
f_1(\beta\omega) & = & \left[ \cosh(\beta\omega) - \cos(\pi/g)
    \right] \left(\beta\omega\right)^{1/g-2} \\
& \times & \left|
    \Gamma\left(1-\frac{1}{2g}-\frac{i\beta\omega}{2\pi}\right) \right|^2
     e^{\beta\omega/2}\nonumber.
\end{eqnarray}
Additionally, we can determine the ratio of the second and fourth
order tunnelling. Using (\ref{eq:final_2nd_order}) and
(\ref{eq:final_4th_order}), one obtains
\begin{equation}         \label{n2n4}
    \frac{n_2(\omega)}{n_4(\omega)} = \frac{2g}{\pi}\Gamma(1-1/g)
    \sin(\pi/g) \frac{k_2}{k_4} \left(\frac{aeV}{\hbar
    v}\right)^{1/g}
    \frac{e\omega}{I\hbar}.
\end{equation}
In order to access information about the nature of correlations
inside the host it is sufficient to compare the experimental data
to the results in the lowest order in tunnelling, as the latter
constitute the dominant contributions and are known in great
detail. However, they are the more pronounced the closer to $E_F$
$j_>(\omega)$ is measured. In these regions the influence of both
$n_4$ and $n_T$ becomes important. Therefore, from the
experimental point of view it is very important to possess
detailed information about the energy window where only $n_2$
yields the major contribution. Having established the relations
(\ref{n2n4}) and (\ref{n2nT}) between all three principal
contributions to the high-energy tails we are now in a position to
accomplish this task.

\begin{figure}[t]
\centering
\includegraphics[width=6cm,angle=270]{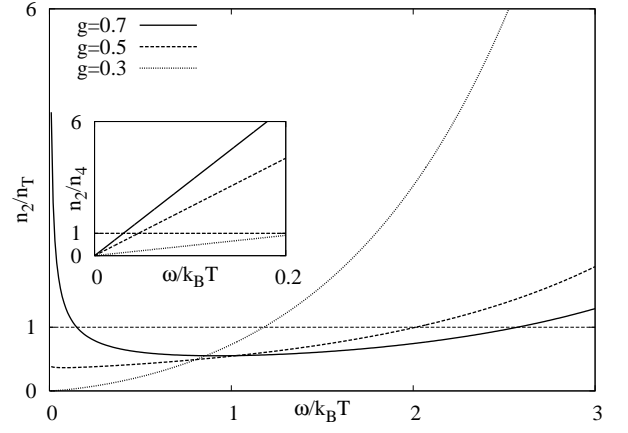}
\caption{Ratio of the secondary current to thermal broadening.
\emph{In the
    inset:} Ratio of the second order to the fourth order
    contribution}
\label{fig:ratio}
\end{figure}

The lower threshold of experimentally measurable currents is
expected to be about $1$nA. Experimentally applicable electric
fields are in the range of $1$V/$\mu$m and the interface width is
of the order of magnitude of $1$nm. Taking into account the
amplification factor due to the sharpness of the emitter tip of
$\sim 10^3$ we estimate the cut-off parameter $V$ to be around
$1$V. Assuming the experiment to be performed at $T=1K$, the
resulting graphs for the ratios of $n_2(\omega)$ to $n_T(\omega)$
and $n_2(\omega)$ to $n_4(\omega)$ are presented in
Fig.~\ref{fig:ratio}. In order to compare the energy scales, we
note that at a temperature of $1$K and a current of $1$nA, $k_B T
\approx 20 I\hbar/e$.
With our choice of parameters, we arrive at the following values
for the lower thresholds for $g \approx 0.5$:
    $n_2 > n_T  \text{ for }  \omega > 2 k_B T$,  and
    $n_2 > n_4  \text{ for }  \omega > 0.05 k_B T$,
so that at least at $1$K the processes of higher order in
tunnelling can safely be neglected. The lower threshold for
$j_>(\omega)$ is then of the order of temperature. The upper
threshold for the second order in tunnelling is given by $\omega =
eV$ as pointed out in Section \ref{principalpart}.

\section{Conclusions}
We have investigated field emission from strongly correlated
emitters with special focus on the high-energy tails of the energy
resolved current. Using Keldysh diagram technique, we could
isolate the relevant diagrams in fourth order in the tunnelling
amplitude. Modelling the interaction in the host wire by a
Luttinger liquid with an interaction parameter $g$ we could
reproduce the already known leading asymptotics of the second
order result, $j_> \sim \omega^{1/g-2}$ by an alternative
approach. Using this method we succeeded in identifying the
leading behaviour of the high-energy tails in fourth order in the
tunnelling amplitude, which turns out to be given by
$\omega^{1/g-3}$. Furthermore, we addressed the question of the
observability of the secondary current by comparing the second
order in tunnelling amplitude contribution with the broadening of
the Fermi edge at finite temperatures as well as with the fourth
order contribution. With typical parameter values at hand we made
an estimation of the energy window where only tunnelling processes
of second order give dominant contributions to the particle energy
distribution above the Fermi edge.

\section*{Acknowledgments}
The authors would like to thank A.~O.~Gogolin and H.~Grabert for
many inspiring discussions. This work was supported by the
Landesstiftung Baden--W\"urttemberg gGmbH (Germany) and by the EC
network DIENOW.

\bibliography{Bibliography}

\end{document}

%% file: aufbau.pstex_t
\begin{picture}(0,0)%
\includegraphics{aufbau.pstex}%
\end{picture}%
\setlength{\unitlength}{3947sp}%
\begingroup\makeatletter\ifx\SetFigFont\undefined%
\gdef\SetFigFont#1#2#3#4#5{%
  \reset@font\fontsize{#1}{#2pt}%
  \fontfamily{#3}\fontseries{#4}\fontshape{#5}%
  \selectfont}%
\fi\endgroup%
\begin{picture}(2724,2095)(4189,-3851)
\put(4201,-2836){\makebox(0,0)[lb]{\smash{\SetFigFont{8}{9.6}{\familydefault}{\mddefault}{\updefault}{\color[rgb]{0,0,0}$\mu_L=0$}%
}}}
\put(6901,-3286){\makebox(0,0)[rb]{\smash{\SetFigFont{8}{9.6}{\familydefault}{\mddefault}{\updefault}{\color[rgb]{0,0,0}$\mu_R=-eV$}%
}}}
\put(5551,-3811){\makebox(0,0)[b]{\smash{\SetFigFont{8}{9.6}{\familydefault}{\mddefault}{\updefault}{\color[rgb]{0,0,0}$x=0$}%
}}}
\put(4201,-2086){\makebox(0,0)[lb]{\smash{\SetFigFont{8}{9.6}{\familydefault}{\mddefault}{\updefault}{\color[rgb]{0,0,0}$\psi$}%
}}}
\put(5551,-2086){\makebox(0,0)[b]{\smash{\SetFigFont{8}{9.6}{\familydefault}{\mddefault}{\updefault}{\color[rgb]{0,0,0}$\gamma$}%
}}}
\put(6901,-2086){\makebox(0,0)[rb]{\smash{\SetFigFont{8}{9.6}{\familydefault}{\mddefault}{\updefault}{\color[rgb]{0,0,0}$c$}%
}}}
\put(4201,-1861){\makebox(0,0)[lb]{\smash{\SetFigFont{12}{14.4}{\familydefault}{\mddefault}{\updefault}{\color[rgb]{0,0,0}host}%
}}}
\put(6901,-1861){\makebox(0,0)[rb]{\smash{\SetFigFont{12}{14.4}{\familydefault}{\mddefault}{\updefault}{\color[rgb]{0,0,0}lead}%
}}}
\put(5851,-3136){\makebox(0,0)[lb]{\smash{\SetFigFont{8}{9.6}{\familydefault}{\mddefault}{\updefault}{\color[rgb]{0,0,0}$eV$}%
}}}
\end{picture}